\documentclass[prb,twocolumn]{revtex4}

\usepackage{amsmath}
\usepackage{amssymb}
\usepackage{amsthm}
\usepackage{amsfonts}
\usepackage{algorithmic}
\usepackage{enumerate}
\usepackage{latexsym}
\usepackage[dvips]{graphicx}

\begin{document}

\title{ Non-adiabatic Arbitrary Geometric Phase Gate in 2-qubit NMR Model}
\author{Yu Tong}
\affiliation{Department of Physics, Fudan University, Shanghai 200433, China}
\author{Ruibao Tao\footnote{To whom correspondence should be addressed. Email:
rbtao@fudan.edu.cn}}
\affiliation{Chinese Center of Advanced Science and Technology (World Laboratory) , \\
P. O. Box 8730 Beijing 100080, China\\
Department of Physics, Fudan University, Shanghai 200433, China}
\date{\today }

\begin{abstract}
We study a 2-qubit nuclear spin system for realizing an arbitrary
geometric quantum phase gate by means of non-adiabatic operation.
A single magnetic pulse with multi harmonic frequencies is applied
to manipulate the quantum states of 2-qubit instantly. Using
resonant transition approximation, the time dependent Hamiltonian
of two nuclear spins can be solved analytically. The time
evolution of the wave function is obtained without adiabatic
approximation. The parameters of magnetic pulse, such as the
frequency, amplitude, phase of each harmonic part as well as the
time duration of the pulse, are determined for achieving an
arbitrary non-adiabatic geometric phase gate. The derivation of
non-adiabatic geometric controlled phase gates and A-A phase are
also addressed.
\end{abstract}

\pacs{03.65.Vf, 76.60.-k, 03.67.Lx}
\maketitle

Taking the advantage of quantum superposition and entanglement, a powerful
new computational algorithms, such as factoring large number\cite{Shor} and
searching an unsorted database\cite{Grover}, has been created. Further study
of quantum information processing and realization of quantum computer has
attracted large numbers of theoretical and experimental scientists. For
implementation of quantum computing, the ability to perform any single
quantum bit (qubit) rotations (SU(2) transformation) and 2-qubit controlled
operations, i.e. controlled not (CNOT) or controlled phase gate (CPG), is
the key requirement \cite{DPD, MAN, SLloyd, Barenco, sl, Turchette, Deutsch}%
. Moreover, each individual qubit should have a long decoherence time to
prevent the deformation of quantum information in qubits for enough long
time. In order to implement a controlled quantum gate (CNOT or CPG), the
"interaction" between qubits is necessary. In this letter, we are interested
in the phase gate of two nuclear spin system. In general, the phase shift of
the quantum state stored in qubits after a cyclic time evolution includes
both the dynamical and geometric part. In some condition, the phase can be
dependent only on the path of the state evolution, not the dynamics at all.
In the case, the relevant transformation of the quantum state is called as a
geometric phase gate \cite{MVB,Shapere,Wilczek} which has been discussed and
implemented in many schemes including trapped ions \cite%
{Schmidt-Kaler,Liebfried}, cavity QED \cite%
{Rauschebeutel,Zheng,Garcia-Maraver} and nuclear magnetic resonance \cite%
{Jonathan,Stadelhofer,Bin}. However, the adiabatic implementation of
geometric phase gate may create some considerable errors \cite{shi}.
Non-adiabatic implementation of quantum information procession is required
in the practice since any quantum computing is requested as fast as
possible. Hence, the theoretical study of non-adiabatic process in quantum
information procession is important and has attracted many scientists. In
this paper, we focus on the study of the implementation of non-adiabatic
geometric phase gate \cite{aharonov}.

Since the nuclear spins are well isolated from the environment, their
decoherence time are far longer. Meanwhile, nuclei with spin 1/2 are the
natural as qubits in quantum information processing. In terms of nuclear
magnetic resonance (NMR), the system of nuclear spins has been a good
candidate to demonstrate the quantum computational algorithms \cite%
{gkk,cvlzll,Vandersypen}. Over the recent several years, many kinds of
complex quantum information processing have been realized by using NMR,
ranging from two to seven qubits in size, in liquid samples \cite{Jones,
Nielsen, Somaroo, Knill, lmkVandersypen} and in solid state samples \cite%
{zhang,Leskowitz,MNDL}. The physical manipulation of nuclear spin state can
be realized by sequences of magnetic pulses with some active resonant
frequencies.

In this paper, we study the 2-qubit system that is composed by two weak
coupled nuclear spins in an external magnetic field. We present details of
the deduction of the time evolution of the quantum state of two nuclear
spins manipulated by an external transverse magnetic pulse with some
selective resonance frequencies, and show how to achieve an arbitrary
non-adiabatic geometric quantum phase gate. Then, as a results, the
controlled geometric phase gate as well as A-A phase are easily obtained.

We first recall the definition of the quantum phase gate (QPG). Denote $T$
as an unitary transformation evolving the input quantum state $|\Psi
(0)\rangle $ to output one $|\Psi (t)\rangle =T|\Psi (0)\rangle $, where $T$
is a $2^{n}\times 2^{n}$ matrix ($n$ the number of qubits). The $T$ defines
a quantum gate. The gate is called as a phase gate if $T_{ij}=e^{i\phi
_{i}}\delta _{ij},i,j=1,2,\cdot \cdot \cdot ,2^{n},$ where $\{\phi
_{i}:i=1,2,....\}$ are the phases that are dependent on the evolution path
in parameter space as well as the dynamics. Thereby each phase $\phi _{i}$
includes both the geometric and dynamical part in general. But, the
dynamical part of $\phi _{i}$ can be vanished in some condition. In the
case, the phase becomes pure geometric and the transformation $T$ is called
as a geometric phase gate (GPG). If the evolution of the state is
instantaneous, $T$ the non-adiabatic GPG. In our present work, we study the
implementation of non-adiabatic GPG.

The Hamiltonian of two nuclear spins with a weak Heisenberg type interaction
in a constant longitudinal magnetic field along z direction is
\begin{eqnarray}
H^{(2)} &=&H_{z}^{(2)}+H_{xy}^{(2)},  \label{1} \\
H_{z}^{(2)} &=&-\frac{1}{2}(\omega _{1}\sigma _{1}^{z}+\omega _{2}\sigma
_{2}^{z}+J\sigma _{1}^{z}\sigma _{2}^{z}),  \label{2} \\
H_{xy}^{(2)} &=&-\frac{1}{2}(J\sigma _{1}^{x}\sigma _{2}^{x}+J\sigma
_{1}^{y}\sigma _{2}^{y}),  \notag
\end{eqnarray}%
where isotropic coupling is assumed, $\omega _{1}$ and $\omega _{2}$ are the
Larmor frequencies of two nuclear spins, $J$ the coupling constant, $%
\{\sigma _{i}^{x},\sigma _{i}^{y},\sigma _{i}^{z}:i=1,2\}$ the Pauli
matrices, and $\hbar =1$. In the experiments, two different nuclear spins
are selected, $\omega _{1}\neq \omega _{2}$, we assume $\omega _{1}>\omega
_{2}$ and the longitudinal constant magnetic field is in the order of $1THz,$
so $\omega _{1},\omega _{2}$ are much large than $J$ and $\eta =J/(\omega
_{1}-\omega _{2})<<$ $1$. The experiment numbers of these parameters will be
given in late$.$ $H_{xy}^{(2)}$ is non-diagonal in $\sigma _{z}$
representation and gives the quantum fluctuation which yields a correction
of order $\eta ^{2}$, hence it can be ignored. Thus, the Ising part $%
H_{z}^{(2)}$ of the Hamiltonian is a well precise approximation. $%
H^{(2)}(\simeq H_{z}^{(2)})$ has four eigenstates: $\{|00\rangle ,|01\rangle
,|10\rangle ,|11\rangle \}$, where $0$ denotes the spin up and $1$ the spin
down. These states correspond to following 4 eigenvalues respectively:%
\begin{eqnarray*}
\epsilon _{1} &=&-(\omega _{1}+\omega _{2}+J)/2,\epsilon _{2}=-(\omega
_{1}-\omega _{2}-J)/2, \\
\epsilon _{3} &=&(\omega _{1}-\omega _{2}+J)/2,\epsilon _{4}=(\omega
_{1}+\omega _{2}-J)/2
\end{eqnarray*}

In this paper, we want to find out an appropriate magnetic pulse to achieve
arbitrary non-adiabatic geometric QPG. Therefor a rectangular transversal
magnetic pulse with $4$-frequency ($\Omega _{1}$, $\Omega _{2}$, $\Omega _{3}
$, $\Omega _{4}$) is applied to the system. The Hamiltonian becomes
\begin{equation}
H_{total}=H_{z}^{(2)}+H_{pulse}(t),  \label{3}
\end{equation}%
where
\begin{eqnarray}
H_{pulse}(t) &=&-\frac{1}{4}\sum_{i=1}^{2}\underset{k=1}{\overset{4}{\sum }}%
h_{ik}(t),  \label{4} \\
h_{ik}(t) &=&h_{ik}(f_{k}(t)\sigma _{i}^{+}+f_{k}^{\ast }(t)\sigma _{i}^{-}).
\label{5}
\end{eqnarray}%
$H_{pulse}(t)$ is the external magnetic pulse \cite{MNDL} where $%
f_{k}(t)=e^{i(\Omega _{k}t+\Phi _{k})}$, $\sigma _{i}^{\pm }=\sigma
_{i}^{x}\pm i\sigma _{i}^{y}$, $\sigma _{i}^{x}$ and $\sigma _{i}^{y}$ are
Pauli operators. $h_{1k}=\gamma _{1}\widetilde{h}_{k}$ and $h_{2k}=\gamma
_{2}\widetilde{h}_{k}$ and $\widetilde{h}_{k}$ the amplitudes of magnetic
pulse. $\gamma _{1}$ and $\gamma _{2}$ are the gyromagnetic ratio of two
different nuclear spins. It is assumed that the magnetic pulse is applied in
the duration $[0,\tau ]:$%
\begin{eqnarray*}
\widetilde{h}_{k} &\neq &0,t\in \lbrack 0,\tau ] \\
\widetilde{h}_{k} &=&0,t\notin \lbrack 0,\tau ],k=1,2,3,4.
\end{eqnarray*}
The phases $\{\Phi _{k}\}$ and duration $\tau $ of the magnetic pulse will
be designed below in detail. Set
\begin{eqnarray}
\Omega _{1} &=&\epsilon _{3}-\epsilon _{1}=\omega _{1}+J,\longrightarrow
|00\rangle \longleftrightarrow |10\rangle   \label{6} \\
\Omega _{2} &=&\epsilon _{4}-\epsilon _{3}=\omega _{2}-J,\longrightarrow
|10\rangle \longleftrightarrow |11\rangle ,  \label{7} \\
\Omega _{3} &=&\epsilon _{2}-\epsilon _{1}=\omega _{2}+J,\longrightarrow
|00\rangle \longleftrightarrow |01\rangle ,  \label{8} \\
\Omega _{4} &=&\epsilon _{4}-\epsilon _{2}=\omega _{1}-J,\longrightarrow
|01\rangle \longleftrightarrow |11\rangle .  \label{9}
\end{eqnarray}%
Corresponding transitions are shown in figure. The Schr\"{o}dinger equation
is
\begin{equation*}
i\hbar \partial |\Psi (t)\rangle )/\partial t=H_{total}(t)|\Psi (t)\rangle ,
\end{equation*}%
The solution can be expended as follows%
\begin{equation}
|\Psi (t)\rangle =\sum\limits_{j}c_{j}(t)|m_{j}\rangle \text{ \ \ }%
(j=1,2,3,4)  \label{10}
\end{equation}%
where $|m_{1}\rangle ,|m_{2}\rangle ,|m_{3}\rangle ,|m_{4}\rangle $
respectively corresponds to the state $|00\rangle ,|01\rangle ,|10\rangle
,|11\rangle .$ When the system is initially at the state $|\Psi (0)\rangle
=|00\rangle $, the pulse with 4 frequencies $\{\Omega _{i}:$ $i=1,2,3,4\}$
leads a coherent evolution of four states $|00\rangle $,$|01\rangle $,$%
|10\rangle $,$|11\rangle $. Since those transitions satisfying
resonant condition are dominating, we can ignore all other minor
detuned transitions. This is so-called resonant transition
approximation (RTA). In RTA, the states transferred each other by
the magnetic pulse are called as the active states, the rest the
inactive. When we apply an magnetic pulse with 4 frequencies (see
Fig.1), in RTA the Hamiltonian $H_{total}(t)$ can be well
approximated by following $\widetilde{H}(t)$,
\begin{equation}
\widetilde{H}(t)=%
\begin{pmatrix}
\epsilon _{1} & -\frac{h_{3}}{2}f_{3}(t) & -\frac{h_{1}}{2}f_{1}(t) & 0 \\
-\frac{h_{3}}{2}f_{3}^{\ast }(t) & \epsilon _{2} & 0 & -\frac{h_{4}}{2}%
f_{4}(t) \\
-\frac{h_{1}}{2}f_{1}^{\ast }(t) & 0 & \epsilon _{3} & -\frac{h_{2}}{2}%
f_{2}(t) \\
0 & -\frac{h_{4}}{2}f_{4}^{\ast }(t) & -\frac{h_{2}}{2}f_{2}^{\ast }(t) &
\epsilon _{4}%
\end{pmatrix}%
,  \label{11}
\end{equation}%
where $h_{1}=h_{11}$, $h_{2}=h_{22}$, $h_{3}=h_{13}$, $h_{4}=h_{24}$ and $%
f_{k}^{\ast }(t)$ is the complex conjugate of $f_{k}(t).$

\begin{figure}
\includegraphics[scale=0.42]{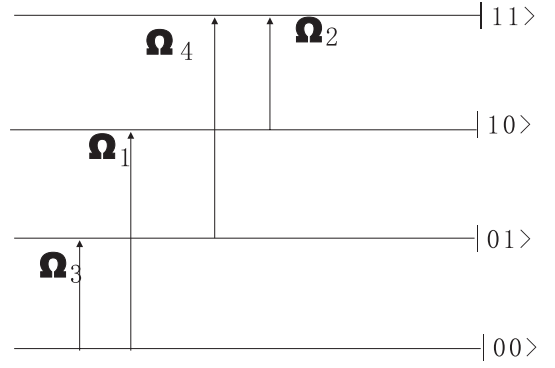}
\caption{The sketch of the energy levels of the Hamiltonian
$H_{total}$ and two multi-photon transition schemes}
\end{figure}

Now we want to find out the exact solution of Hamiltonian without adiabatic
approximation. We do an unitary transformation $U(t)$ as follows%
\begin{equation}
|\Psi (t)\rangle =U^{+}(t)|\psi (t)\rangle .  \label{12}
\end{equation}%
From Schr\"{o}dinger equation $i\hbar \partial (|\Psi (t\mathbf{)\rangle )}%
/\partial t=\widetilde{H}(t)|\Psi (t)\rangle $, we obtain an equation of $%
|\psi (t)\rangle :$
\begin{equation}
i\hbar \partial (|\psi (t)\rangle )/\partial t=H_{rot}(t)|\psi (t)\rangle ,
\label{13}
\end{equation}%
where
\begin{equation}
H_{rot}(t)=U(t)\widetilde{H}(t)U^{+}(t)-iU(t)\partial U^{+}(t)/\partial
t.\qquad   \label{14}
\end{equation}%
It is assumed that the matrix $U(t)$ is diagonal: $U_{ij}(t)=\delta
_{ij}e^{i(\varphi _{i}t+\theta _{i})}$ where $\{\varphi _{i},\theta _{i}\}$
are some parameters to be determined$.$ When the parameters $\{\varphi
_{i},\theta _{i}\}$ satisfy the following relations:
\begin{eqnarray}
\Omega _{1} &=&\varphi _{3}-\varphi _{1},\Phi _{1}=\theta _{3}-\theta _{1}
\label{15} \\
\Omega _{2} &=&\varphi _{4}-\varphi _{3},\Phi _{2}=\theta _{4}-\theta _{3},
\label{16} \\
\Omega _{3} &=&\varphi _{2}-\varphi _{1},\Phi _{3}=\theta _{2}-\theta _{1}
\label{17} \\
\Omega _{4} &=&\varphi _{4}-\varphi _{2},\Phi _{4}=\theta _{4}-\theta _{2}
\label{18}
\end{eqnarray}%
It can be easily found that%
\begin{equation}
H_{rot}(t)=%
\begin{pmatrix}
\epsilon _{1}-\varphi _{1} & -\frac{h_{3}}{2} & -\frac{h_{1}}{2} & 0 \\
-\frac{h_{3}}{2} & \epsilon _{2}-\varphi _{2} & 0 & -\frac{h_{4}}{2} \\
-\frac{h_{1}}{2} & 0 & \epsilon _{3}-\varphi _{3} & -\frac{h_{2}}{2} \\
0 & -\frac{h_{4}}{2} & -\frac{h_{2}}{2} & \epsilon _{4}-\varphi _{4}%
\end{pmatrix}%
.  \label{19}
\end{equation}%
which is time independent. We simply denote $H_{rot}(t)$ as $H_{rot}.$ Due
to $\Omega _{1}=\epsilon _{3}-\epsilon _{1},$ $\Omega _{2}=$ $\epsilon
_{4}-\epsilon _{3},$ $\Omega _{3}=\epsilon _{2}-\epsilon _{1},$ $\Omega
_{4}=\epsilon _{4}-\epsilon _{2},$ then considering the relations of
(15)-(18), we get%
\begin{eqnarray*}
\epsilon _{2}-\varphi _{2} &=&\Omega _{3}+\epsilon _{1}-\varphi
_{2}=\epsilon _{1}-\varphi _{1}, \\
\epsilon _{3}-\varphi _{3} &=&\Omega _{1}+\epsilon _{1}-\varphi
_{3}=\epsilon _{1}-\varphi _{1}, \\
\epsilon _{4}-\varphi _{4} &=&\Omega _{3}+\Omega _{4}+\epsilon _{1}-\varphi
_{4}=\epsilon _{1}-\varphi _{1}.
\end{eqnarray*}%
Setting $\varphi _{1}=\epsilon _{1},$ equation (19) becomes%
\begin{equation}
H_{rot}(t)=H_{rot}=%
\begin{pmatrix}
0 & -\frac{h_{3}}{2} & -\frac{h_{1}}{2} & 0 \\
-\frac{h_{3}}{2} & 0 & 0 & -\frac{h_{4}}{2} \\
-\frac{h_{1}}{2} & 0 & 0 & -\frac{h_{2}}{2} \\
0 & -\frac{h_{4}}{2} & -\frac{h_{2}}{2} & 0%
\end{pmatrix}%
,  \label{20}
\end{equation}%
From equations (15)-(18), we have%
\begin{eqnarray*}
\varphi _{2} &=&\Omega _{3}+\varphi _{1},\varphi _{3}=\Omega _{1}+\varphi
_{1}, \\
\varphi _{4} &=&\Omega _{1}+\Omega _{2}-\varphi _{1}, \\
\theta _{2} &=&\Phi _{3}+\Phi _{1},\theta _{3}=\Phi _{1}+\theta _{1}, \\
\theta _{4} &=&\Phi _{1}+\Phi _{2}-\theta _{1}.
\end{eqnarray*}%
Four eigenvalues of $H_{rot}$ can be found as following
\begin{eqnarray*}
E_{1} &=&\frac{-\sqrt{2}}{4}\sqrt{A+\sqrt{B+C}}\text{,} \\
E_{2} &=&\frac{\sqrt{2}}{4}\sqrt{A+\sqrt{B+C}}\text{,} \\
E_{3} &=&\frac{-\sqrt{2}}{4}\sqrt{A-\sqrt{B+C}}\text{,} \\
E_{4} &=&\frac{\sqrt{2}}{4}\sqrt{A-\sqrt{B+C}}\text{,}
\end{eqnarray*}%
where%
\begin{eqnarray*}
A &=&\sum_{i=1}^{4}h_{i}^{2}\text{, \ \ \ \ }B=\sum_{i=1}^{4}h_{i}^{4}, \\
C &=&8h_{1}h_{2}h_{3}h_{4}+2h_{1}^{2}h_{2}^{2}+2h_{1}^{2}h_{3}^{2} \\
&&-2h_{1}^{2}h_{4}^{2}-2h_{2}^{2}h_{3}^{2}+2h_{2}^{2}h_{4}^{2}+2h_{3}^{2}h_{4}^{2}
\end{eqnarray*}%
we further set $h_{1}=h_{4}$ and $h_{2}=h_{3}$. It results%
\begin{eqnarray}
E_{1} &=&-\frac{1}{2}(h_{1}+h_{2})\text{, }E_{2}=\frac{1}{2}(h_{1}+h_{2})%
\text{,}  \label{21} \\
E_{3} &=&-\frac{1}{2}(h_{1}-h_{2})\text{, }E_{4}=\frac{1}{2}(h_{1}-h_{2})%
\text{.}  \label{22}
\end{eqnarray}%
$\allowbreak $Four corresponding eigenfunctions are $|\psi _{i}\rangle
=\sum_{j=1}^{4}C_{ij}|m_{j}\rangle ,i=1,2,3,4.$ The coefficients matrix $C$
and its inverse $C^{-1}$ are%
\begin{equation}
C=C^{-1}=\frac{1}{2}\left(
\begin{array}{cccc}
1 & 1 & 1 & 1 \\
1 & -1 & -1 & 1 \\
1 & -1 & 1 & -1 \\
1 & 1 & -1 & -1%
\end{array}%
\right) .\text{ }  \label{23}
\end{equation}%
$\allowbreak |m_{i}\rangle =\sum_{j=1}^{4}C_{ij}^{-1}|\mathbf{\psi }%
_{j}\rangle ,i=1,2,3,4.$. If the system is initially at the one of the
eigenstates: $|\Psi ^{(i)}(0)\rangle =|m_{i}\rangle ,$ $(i=1,2,3,4)$, the
state of system at time $t$ will be%
\begin{eqnarray}
|\Psi ^{(i)}(t)\rangle  &=&U^{+}(t)\exp [-iH_{rot}t]\left\vert
m_{i}\right\rangle   \notag \\
&=&\sum_{j=1}^{4}C_{ij}^{-1}U^{+}(t)\exp [-iH_{rot}t]|\mathbf{\psi }_{j}%
\mathbf{\rangle }  \notag \\
&=&\sum_{j,k=1}^{4}C_{ij}^{-1}|m_{k}\rangle \langle m_{k}|U^{+}(t)|\mathbf{%
\psi }_{j}\mathbf{\rangle }\exp (-iE_{j}t)  \notag \\
&=&\sum_{j,k=1}^{4}C_{ij}^{-1}e^{-i(\varphi _{k}t+\theta _{k})}\exp
(-iE_{j}t)C_{jk}|m_{k}\rangle   \notag \\
&=&\sum_{k}c_{k}^{(i)}(t)|m_{k}\rangle ,i=1,2,3,4,  \label{24}
\end{eqnarray}%
where
\begin{eqnarray}
c_{k}^{(i)}(t) &=&\sum_{j=1}^{4}\left( C_{ij}^{-1}\exp
(-iE_{j}t)C_{jk}\right) e^{-i(\varphi _{k}t+\theta _{k})}  \notag \\
&=&\widetilde{c}_{k}^{(i)}(t)e^{-i(\varphi _{k}t+\theta _{k})}  \label{25} \\
\widetilde{c}_{k}^{(i)}(t) &=&\overset{4}{\underset{j=1}{\sum }}%
C_{ij}^{-1}e^{-iE_{j}t}C_{jk}.  \label{26}
\end{eqnarray}%
The initial condition is $|c_{i}^{(i)}(0)|^{2}=1$ and $|c_{k}^{(i)}(0)|^{2}=0
$ $(k\neq i)$.

For a cyclic evolution after time $\tau $,
\begin{equation}
{|{\Psi }}^{(i)}{\mathbf{(}}\tau {\mathbf{)}\rangle }=e^{i\beta _{i}(\tau
)}|\Psi \mathbf{(0)}\rangle ,  \label{27}
\end{equation}%
where the total phase is $\beta _{i}(\tau )=\arg {\langle }\Psi ^{(i)}{%
\mathbf{(}}0{\mathbf{)}|}\Psi ^{(i)}{\mathbf{(}}\tau {\mathbf{)}\rangle }%
=\delta _{D}^{(i)}(\tau )+\delta _{G}^{(i)}(\tau )$. $\delta _{G}^{(i)}(\tau
)$ is the geometric phase. The dynamical phase $\delta _{D}^{(i)}(\tau )$
can be calculated by formula:%
\begin{eqnarray}
\delta _{D}^{(i)}(\tau ) &=&-\int_{0}^{\tau }dt\langle \Psi ^{(i)}\mathbf{(}t%
\mathbf{)}|\widetilde{H}(t)|\Psi ^{(i)}\mathbf{(}t\mathbf{)}\rangle   \notag
\\
&=&-\int_{0}^{\tau }dt\sum\limits_{k,k^{\prime }}c_{k}^{(i)\ast
}(t)c_{k^{\prime }}^{(i)}(t)\langle m_{k}|\widetilde{H}(t)|m_{k^{\prime
}}\rangle   \label{28}
\end{eqnarray}%
In appendix A, we obtain
\begin{eqnarray}
\widetilde{c}_{1}^{(1)}(t) &=&\widetilde{c}_{2}^{(2)}(t)=\widetilde{c}%
_{3}^{(3)}(t)=\widetilde{c}_{4}^{(4)}(t)  \notag \\
&=&\cos \left( \frac{h_{1}}{2}t\right) \cos \left( \frac{h_{2}}{2}t\right)
\label{29} \\
\widetilde{c}_{2}^{(1)}(t) &=&\widetilde{c}_{1}^{(2)}(t)=\widetilde{c}%
_{4}^{(3)}(t)=\widetilde{c}_{3}^{(4)}(t)  \notag \\
&=&i\cos [\frac{h_{1}}{2}t]\sin [\frac{h_{2}}{2}t]  \label{30} \\
\widetilde{c}_{3}^{(1)}(t) &=&\widetilde{c}_{4}^{(2)}(t)=\widetilde{c}%
_{1}^{(3)}(t)=\widetilde{c}_{2}^{(4)}(t)  \notag \\
&=&i\sin [\frac{h_{1}}{2}t]\cos [\frac{h_{2}}{2}t]  \label{31} \\
\widetilde{c}_{4}^{(1)}(t) &=&\widetilde{c}_{3}^{(2)}(t)=\widetilde{c}%
_{2}^{(3)}(t)=\widetilde{c}_{1}^{(4)}(t)  \notag \\
&=&\sin [\frac{h_{1}}{2}t]\sin [\frac{h_{2}}{2}t]  \label{32}
\end{eqnarray}%
When $t=\tau $ and satisfies $h_{1}\tau =2m\pi ,h_{2}\tau =2n\pi ,$ namely
\begin{equation}
\tau =2m\pi /h_{1},h_{2}=\frac{n}{m}h_{1.},m,n=1,2,3,..
\end{equation}%
we have%
\begin{eqnarray*}
|\widetilde{c}_{i}^{(i)}(\tau )| &=&1:i=1,2,3,4, \\
\widetilde{c}_{i}^{(j)}(\tau ) &=&0:i\neq j.
\end{eqnarray*}%
It directly yields
\begin{eqnarray*}
|\Psi ^{(i)}(\tau )\rangle  &=&\sum_{k}e^{-i(\varphi _{k}t+\theta _{k})}%
\widetilde{c}_{k}^{(i)}(t)|m_{k}\rangle  \\
&=&e^{-i(\varphi _{i}\tau +\theta _{i})}|m_{i}\rangle ,i=1,2,3,4.
\end{eqnarray*}%
Equation (33) is the condition to achieve a phase gate. In appendix B, we
give the proof of that the dynamical phases also vanish at same condition
(33):
\begin{equation}
\delta _{D}^{(i)}(\tau )=0,i=1,2,3,4.  \label{34}
\end{equation}%
Hence, for any initial state $|\Psi (0)\rangle
=\sum_{i}c_{i}(0)|m_{i}\rangle $ the wave function $|\Psi (\tau )\rangle $
is
\begin{equation}
|\Psi (\tau \rangle =A\left(
\begin{array}{cccc}
e^{-i\Theta _{1}} & 0 & 0 & 0 \\
0 & e^{-i\Theta _{2}} & 0 & 0 \\
0 & 0 & e^{-i\Theta _{3}} & 0 \\
0 & 0 & 0 & e^{-i\Theta _{4}}%
\end{array}%
\right) |\Psi (0)\rangle .  \label{35}
\end{equation}%
Where $A(=\cos (m\pi /2)\cos (n\pi /2),A=\pm 1)$ only contributes a global
phase. The phases $\{\Theta _{i}=\varphi _{i}\tau +\theta _{i}:i=1,2,3,4\}$
are all geometric. Hence a non-adiabatic geometric phase gate is achieved.

From equations (15) to (18), there are relations $\varphi _{2}=\Omega
_{3}+\varphi _{1},$ $\varphi _{3}=\Omega _{1}+\varphi _{1},$ $\varphi
_{4}=\Omega _{1}+\Omega _{2}-\varphi _{1},\varphi _{1}=\epsilon _{1}.$ It is
similar to have $\theta _{4}=\Phi _{1}+\Phi _{2}-\theta _{1},$ $\theta
_{2}=\Phi _{3}+\Phi _{1},$ $\theta _{3}=\Phi _{1}+\theta _{1}.$ Since all $%
\{\Omega _{i}\}$ and $\epsilon _{1}$ are fixed, all $\{\varphi _{i}\}$ are
changeless. $\{\Phi _{i}\}$ are the phases of four harmonic waves of the
external magnetic pulse and can be freely adjusted. Hence we can use four
free parameters $\{\Phi _{i}\}$ to change the values of $\{\theta
_{i}:i=1,2,3,4\},$then to alter four phases $\Theta _{i}(=\varphi _{i}\tau
+\theta _{i})$ in gate individually. As a result the arbitrary non-adiabatic
geometric phase gate of two qubits is implemented by means of NMR. The
duration time $\tau $ of magnetic pulse is determined by equations (33): $%
\tau =2m\pi /h_{1},h_{2}=\frac{n}{m}h_{1}$ where $h_{1}$ and $h_{2}$ are
related to the amplitude of magnetic pulse and are adjustable. Taking $m=n=1,
$it gives $h_{1}=h_{2}$ and $\tau =2\pi /h_{1}.$

Set $\theta _{1}=-\varphi _{1}\tau $ and $\theta _{2}=-\varphi _{2}\tau $ in
equation (35)$,$ we obtain an arbitrary geometric controlled phase gate
(CPG) by changing the parameters $\{\theta _{3},\theta _{4}\}$:%
\begin{equation*}
T_{CPG}=A\left(
\begin{array}{cccc}
1 & 0 & 0 & 0 \\
0 & 1 & 0 & 0 \\
0 & 0 & e^{-i(\varphi _{3}\tau +\theta _{3})} & 0 \\
0 & 0 & 0 & e^{-i(\varphi _{4}\tau +\theta _{4})}%
\end{array}%
\right) ,A=\pm 1.
\end{equation*}%
Further, if we adjust the parameters $\{\theta _{i}\}$ to satisfy following
conditions%
\begin{equation*}
\varphi _{i}\tau +\theta _{i}=\varphi _{1}\tau +\theta _{1}:i=2,3,4.
\end{equation*}%
The wave function at time $\tau $ is recovered to initial state with a phase
difference only.
\begin{equation*}
|\Psi (\tau )\rangle =\pm e^{-i(\varphi _{1}\tau +\theta _{1})}|\Psi
(0)\rangle .
\end{equation*}%
Where $\varphi _{1}\tau +\theta _{1}$ is just the A-A phase \cite{aharonov}.

Finally, we give brief discussion on the validity of Ising model and our RTA
and some estimation of the time $\tau $ to achieve the geometric arbitrary
phase gate. It is known that the nuclei $^{1}H,^{13}C,^{15}N,^{19}F,^{31}P,$%
etc. have spin $S=1/2.$ Taking the $^{1}H$ and $^{13}C$ as an example \cite%
{Nielsen, Knill}, the Larmor procession frequency of nucleus $^{1}H$ : $%
\omega _{1}\sim 500MHz$ at field $B_{0}=1.87T$. The gyromagnetic ratio of$^{%
\text{ }13}C$ has only $25\%$\ of $^{1}H$ \cite{Inglis}, so $\omega _{2}\sim
125MHz.$ $\omega _{1}-\omega _{2}\sim 375MHz.$ Assume the coupling $J\sim
200Hz$ \cite{Nielsen, Knill}(in the references $J\sim 200Hz$), we have $%
J/(\omega _{1}-\omega _{2})\sim 0.54\times 10^{-6}(<<1).$ Hence the quantum
fluctuation from $H_{xy}^{(2)}$ can be ignored. It shows that our Ising
model Hamiltonian can well describe the physics.

We set the intensity of the external transverse magnetic pulse $\widetilde{h}%
_{1}\sim 10^{-2}T$, it yields $h_{1}\sim 2.8MHz.$ From the equation (33), we
have
\begin{equation*}
\tau =2m\pi /h_{1}\sim 2.2\times m\times 10^{-6}\sec .
\end{equation*}%
Meanwhile, $h_{2}$ should be adjusted to meet the condition $h_{2}=\frac{n}{m%
}h_{1}.$ Since gyromagnetic ratio of $^{13}C$ is $0.25$ of the one of $^{1}H,
$ the amplitude $\widetilde{h}_{2}$ must be four times of $\widetilde{h}_{1}$
($\widetilde{h}_{2}\simeq 4\widetilde{h}_{1}$), then the condition $h_{2}=%
\frac{n}{m}h_{1}$ can be satisfied$.$ We choose the one of solutions, $%
n=1,m=1,$ that yields $h_{2}(=h_{1})\sim 2.8MHz$ and $\tau =2\pi /h_{1}\sim
2.2\times 10^{-6}\sec .$ Due to $\Omega _{1,4}=\omega _{1}\pm J,$ $\Omega
_{2,3}=\omega _{2}\mp J,$ we have $\Omega _{i}\tau \sim 2.5\times
10^{2}-1.0\times 10^{3}>>1$ for all $i.$ Larger $m$ or smaller $h_{1}$
yields more longer $\tau $. Thus, the precision of our RTA can be well
ensured.

The phase decoherence time $\tau _{\phi }$ of nuclear spin is the order of $%
10^{4}$ sec \cite{DPD}, but the operating time $\tau $ of such non-adiabatic
geometric phase gate is only the order of $10^{-6}\sec .$ The ratio $\tau
_{\phi }/\tau (\sim 10^{10}-10^{9})$ in NMR is high.

To summarize, We have studied Ising model of two nuclear spins. A single
tranverse $rf$ magnetic pulse with multi-frequency is applied to manipulate
quantum state simultaneously. By means of RTA, an arbitrary non-adiabatic
geometric phase gate. As a result, the controlled phase gate as well as A-A
phase are also addressed. From estimation of parameters, we believe that our
RTA is a good approximation and the operation time of GPG $\tau $ is order
of $10^{-6}-10^{-5}$ $\sec $. The absence of dynamical contributions to the
phase gate can decrease the error of dynamical fluctuation in quantum
information process. Meanwhile, the non-adiabatic manipulation of quantum
state in practice quantum computing is required that leads the study of
non-adiabatic manipulation to implement the geometric phase gate.

\textbf{Acknowledgement:} We thank Dr. Y. Shi for some helpful discussion.
This work is supported by the China National Natural Science Foundation
(Nos.10234010 and 10674027) and by 973 project (No.2002CB613504) of China
Ministry of Science and Technology.

\begin{center}
\textbf{{Appendix A} }
\end{center}

In this appendix, we would prove $|c_{k}^{(i)}(\tau )|=\delta _{k}^{i}$ when
$h_{1}\tau =2m\pi $ and $h_{2}\tau =2n\pi $ where $m$ and $n$ are integer
numbers. According to eq.(26), $\widetilde{c}_{k}^{(i)}(t)=\overset{4}{%
\underset{j=1}{\sum }}C_{ij}^{-1}e^{-iE_{j}t}C_{jk},$ and eq.(23) of matrix $%
C,$ we get%
\begin{eqnarray*}
\widetilde{c}_{1}^{(1)}(t) &=&\widetilde{c}_{2}^{(2)}(t)=\widetilde{c}%
_{3}^{(3)}(t)=\widetilde{c}_{4}^{(4)}(t) \\
&=&\overset{4}{\underset{j=1}{\sum }}C_{1j}^{-1}e^{-iE_{j}t}C_{j1} \\
&=&\frac{1}{4}\{e^{-iE_{1}t}+e^{-iE_{2}t}+e^{-iE_{3}t}+e^{-iE_{4}t}\} \\
&=&\frac{1}{2}\left( \cos (\frac{h_{1}+h_{2}}{2})+\cos (\frac{h_{1}-h_{2}}{2}%
)\right)  \\
&=&\cos \left( \frac{h_{1}}{2}t\right) \cos \left( \frac{h_{2}}{2}t\right)
\end{eqnarray*}%
\begin{eqnarray*}
\widetilde{c}_{2}^{(1)}(t) &=&\widetilde{c}_{1}^{(2)}(t)=\widetilde{c}%
_{4}^{(3)}(t)=\widetilde{c}_{3}^{(4)}(t) \\
&=&\overset{4}{\underset{j=1}{\sum }}C_{1j}^{-1}e^{-iE_{j}t}C_{j2} \\
&=&\frac{1}{4}[e^{-iE_{1}t}-e^{-iE_{2}t}-e^{-iE_{3}t}+e^{-iE_{4}t}] \\
&=&\frac{i}{2}[\sin \frac{(h_{1}+h_{2})t}{2}-\sin \frac{(h_{1}-h_{2})t}{2}]
\\
&=&i\cos [\frac{h_{1}}{2}t]\sin [\frac{h_{2}}{2}t]
\end{eqnarray*}%
It is similar to obtain following results:%
\begin{eqnarray*}
\widetilde{c}_{2}^{(1)}(t) &=&\widetilde{c}_{1}^{(2)}(t)=\widetilde{c}%
_{4}^{(3)}(t)=\widetilde{c}_{3}^{(4)}(t) \\
&=&i\cos [\frac{h_{1}}{2}t]\sin [\frac{h_{2}}{2}t] \\
\widetilde{c}_{3}^{(1)}(t) &=&\widetilde{c}_{4}^{(2)}(t)=\widetilde{c}%
_{1}^{(3)}(t)=\widetilde{c}_{2}^{(4)}(t) \\
&=&i\sin [\frac{h_{1}}{2}t]\cos [\frac{h_{2}}{2}t] \\
\widetilde{c}_{4}^{(1)}(t) &=&\widetilde{c}_{3}^{(2)}(t)=\widetilde{c}%
_{2}^{(3)}(t)=\widetilde{c}_{1}^{(4)}(t) \\
&=&\sin [\frac{h_{1}}{2}t]\sin [\frac{h_{2}}{2}t]
\end{eqnarray*}%
When $t=\tau ,h_{1}\tau =2m\pi $ and $h_{2}\tau =2n\pi $ ($m,n:$ integer
numbers), following results are obtained:
\begin{equation*}
|\widetilde{c}_{i}^{(i)}(\tau )|=1;\widetilde{c}_{i}^{(j)}(\tau )=0,i\neq j.
\end{equation*}

\begin{center}
\textbf{{Appendix B} }
\end{center}

In this appendix, we want to prove the equation (28): $\delta
_{D}^{(i)}=0:i=1,2,3,4.$

When $h_{1}=h_{4},h_{2}=h_{3}$, the definition of $\widetilde{H}(t)$ becomes
(see eq.(11))
\begin{equation*}
\widetilde{H}(t)=%
\begin{pmatrix}
\epsilon _{1} & -\frac{h_{2}}{2}f_{3}(t) & -\frac{h_{1}}{2}f_{1}(t) & 0 \\
-\frac{h_{2}}{2}f_{3}^{\ast }(t) & \epsilon _{2} & 0 & -\frac{h_{1}}{2}%
f_{4}(t) \\
-\frac{h_{1}}{2}f_{1}^{\ast }(t) & 0 & \epsilon _{3} & -\frac{h_{2}}{2}%
f_{2}(t) \\
0 & -\frac{h_{1}}{2}f_{4}^{\ast }(t) & -\frac{h_{2}}{2}f_{2}^{\ast }(t) &
\epsilon _{4},%
\end{pmatrix}%
\end{equation*}%
where $f_{k}(t)=e^{i(\Omega _{k}t+\Phi _{k})}.$ According to the definition
of dynamic phase, eq.(28), we have
\begin{align*}
-\delta _{D}^{(i)}& =\int_{0}^{\tau }dt\sum\limits_{k,k^{\prime
}}c_{k}^{(i)\ast }(t)c_{k^{\prime }}^{(i)}(t)\langle m_{k}|\widetilde{H}%
(t)|m_{k^{\prime }}\rangle  \\
& =\int_{0}^{\tau }dt\underset{k=1}{\overset{4}{\sum }}\widetilde{c}%
_{k}^{(i)^{\ast }}(t)\widetilde{c}_{k}^{(i)}(t)\epsilon _{k} \\
& -\{\frac{h_{2}}{2}\widetilde{c}_{2}^{(i)\ast }(t)\widetilde{c}%
_{1}^{(i)}(t)e^{-i(\Omega _{3}t+\Phi _{3})-i(\varphi _{1}t+\theta
_{1})+i(\varphi _{2}t+\theta _{2})} \\
& +\frac{h_{1}}{2}\widetilde{c}_{3}^{(i)\ast }(t)\widetilde{c}%
_{1}^{(i)}(t)e^{-i(\Omega _{1}t+\Phi _{1})-i(\varphi _{1}t+\theta
_{1})+i(\varphi _{3}t+\theta _{3})} \\
& +\frac{h_{1}}{2}\widetilde{c}_{4}^{(i)\ast }(t)\widetilde{c}%
_{2}^{(i)}(t)e^{-i(\Omega _{4}t+\Phi _{4})-i(\varphi _{2}t+\theta
_{2})+i(\varphi _{4}t+\theta _{4})} \\
& +\frac{h_{2}}{2}\widetilde{c}_{4}^{(i)\ast }(t)\widetilde{c}%
_{3}^{(i)}(t)e^{-i(\Omega _{2}t+\Phi _{2})-i(\varphi _{3}t+\theta
_{3})+i(\varphi _{4}t+\theta _{4})} \\
& +h.c.\}
\end{align*}%
Using the relations of eq.(15)$-$(18), above equation becomes%
\begin{align}
-\delta _{D}^{(i)}& =\int_{0}^{\tau }dt\sum\limits_{k,k^{\prime
}}c_{k}^{(i)\ast }(t)c_{k^{\prime }}^{(i)}(t)\langle m_{k}|\widetilde{H}%
(t)|m_{k^{\prime }}\rangle   \notag \\
& =\int_{0}^{\tau }dt\underset{k=1}{\overset{4}{\sum }}\widetilde{c}%
_{k}^{(i)^{\ast }}(t)\widetilde{c}_{k}^{(i)}(t)\epsilon _{k}  \notag \\
& -\{\frac{h_{2}}{2}[\widetilde{c}_{2}^{(i)\ast }(t)\widetilde{c}%
_{1}^{(i)}(t)+\widetilde{c}_{4}^{(i)\ast }(t)\widetilde{c}_{3}^{(i)}(t)]
\notag \\
& +\frac{h_{1}}{2}[\widetilde{c}_{3}^{(i)\ast }(t)\widetilde{c}_{1}^{(i)}(t)+%
\widetilde{c}_{4}^{(i)\ast }(t)\widetilde{c}_{2}^{(i)}(t)]+h.c.\}.  \tag{B1}
\end{align}%
We do following calculations:
\begin{eqnarray*}
&&\int_{0}^{\tau }dt\widetilde{c}_{1}^{(1)^{\ast }}(t)\widetilde{c}%
_{1}^{(1)}(t)\epsilon _{1} \\
&=&\epsilon _{1}\int_{0}^{\tau }dt\cos ^{2}\left( \frac{h_{3}}{2}t\right)
\cos ^{2}\left( \frac{h_{4}}{2}t\right)  \\
&=&\frac{\epsilon _{1}}{4}\int_{0}^{\tau }dt[1+\cos (h_{3}t)][1+\cos
(h_{4}t)] \\
&=&\frac{\epsilon _{1}}{4}\left( \tau +\int_{0}^{\tau }dt\cos (h_{3}t)\cos
(h_{4}t)\right)  \\
&=&\frac{\epsilon _{1}}{4}\left( \tau +\frac{1}{2}\int_{0}^{\tau }dt(\cos
(h_{3}+h_{4})t+\cos (h_{3}-h_{4})t\right)  \\
&=&\frac{\epsilon _{1}}{4}\tau .
\end{eqnarray*}%
\begin{eqnarray*}
&&\int_{0}^{\tau }dt\widetilde{c}_{2}^{(1)^{\ast }}(t)\widetilde{c}%
_{2}^{(1)}(t)\epsilon _{2} \\
&=&\epsilon _{2}\int_{0}^{\tau }dt\cos ^{2}\left( \frac{h_{1}}{2}t\right)
\sin ^{2}\left( \frac{h_{2}}{2}t\right)  \\
&=&\frac{\epsilon _{2}}{4}\int_{0}^{\tau }dt[1+\cos (h_{1}t)][1-\cos
(h_{2}t)] \\
&=&\frac{\epsilon _{2}}{4}\left( \tau -\int_{0}^{\tau }dt\cos (h_{1}t)\cos
(h_{2}t)\right) =\frac{\epsilon _{2}}{4}\tau .
\end{eqnarray*}%
It is similar to have%
\begin{eqnarray*}
&&\int_{0}^{\tau }dt\widetilde{c}_{3}^{(1)^{\ast }}(t)\widetilde{c}%
_{3}^{(1)}(t)\epsilon _{2} \\
&=&\epsilon _{3}\int_{0}^{\tau }dt\sin ^{2}\left( \frac{h_{1}}{2}t\right)
\cos ^{2}\left( \frac{h_{2}}{2}t\right) =\frac{\epsilon _{3}}{4}\tau , \\
&&\int_{0}^{\tau }dt\widetilde{c}_{4}^{(1)^{\ast }}(t)\widetilde{c}%
_{4}^{(1)}(t)\epsilon _{4} \\
&=&\epsilon _{4}\int_{0}^{\tau }dt\sin \left( \frac{h_{1}}{2}t\right) \sin
\left( \frac{h_{2}}{2}t\right) =\frac{\epsilon _{4}}{4}\tau .
\end{eqnarray*}%
Input $\epsilon _{1}=-(\omega _{1}+\omega _{2}+J)/2$, $\epsilon
_{2}=-(\omega _{1}-\omega _{2}-J)/2$, $\epsilon _{3}=(\omega _{1}-\omega
_{2}+J)/2$, and $\epsilon _{4}=(\omega _{1}+\omega _{2}-J)/2$ to above
equations, we get%
\begin{equation}
\int_{0}^{\tau }dt\underset{k=1}{\overset{4}{\sum }}\widetilde{c}%
_{k}^{(1)^{\ast }}(t)\widetilde{c}_{k}^{(1)}(t)\epsilon _{k}=\frac{\tau }{4}%
\left( \underset{k=1}{\overset{4}{\sum }}\epsilon _{k}\right) =0.  \tag{B2}
\end{equation}%
Moreover
\begin{align}
& \frac{h_{1}}{2}\int_{0}^{\tau }dt[\widetilde{c}_{3}^{(1)\ast }(t)%
\widetilde{c}_{1}^{(1)}(t)+\widetilde{c}_{4}^{(1)\ast }(t)\widetilde{c}%
_{2}^{(1)}(t)]  \notag \\
& =\frac{h_{1}}{2}\int_{0}^{\tau }dt\{-i\sin [\frac{h_{1}}{2}t]\cos [\frac{%
h_{2}}{2}t]\cos [\frac{h_{2}}{2}t]\cos [\frac{h_{1}}{2}t]  \notag \\
& +i\sin [\frac{h_{1}}{2}t]\sin [\frac{h_{2}}{2}t]\sin [\frac{h_{2}}{2}%
t]\cos [\frac{h_{1}}{2}t]\}  \notag \\
& =-\frac{h_{1}}{4}\int_{0}^{\tau }dt\{-i\sin [h_{1}t]\cos [h_{2}t]\}=0
\tag{B3}
\end{align}%
\begin{align}
& \frac{h_{2}}{2}\int_{0}^{\tau }dt[\widetilde{c}_{4}^{(1)\ast }(t)%
\widetilde{c}_{3}^{(1)}(t)+\widetilde{c}_{2}^{(1)\ast }(t)\widetilde{c}%
_{1}^{(1)}(t)]  \notag \\
& =\frac{h_{2}}{2}\int_{0}^{\tau }dt\{i\sin [\frac{h_{1}}{2}t]\sin [\frac{%
h_{2}}{2}t]\sin [\frac{h_{1}}{2}t]\cos [\frac{h_{2}}{2}t]  \notag \\
& \ \ -i\cos [\frac{h_{1}}{2}t]\sin [\frac{h_{2}}{2}t]\cos [\frac{h_{1}}{2}%
t]\cos [\frac{h_{2}}{2}t]\}  \notag \\
& =\frac{h_{2}}{2}\int_{0}^{\tau }dt\{i\sin (h_{2}t)\cos (h_{1}t)\}=0
\tag{B4}
\end{align}%
Inserting results of (B2-B4) to (B1), we obtain $\delta _{D}^{(1)}(\tau )=0.$
Similar calculations also can show $\delta _{D}^{(2,3,4)}(\tau )=0.$ The
conclusion of $\delta _{D}^{(i)}(\tau )=0$ is proved for all $i$.

\end{document}